\documentclass[aps,prx,reprint,superscriptaddress]{revtex4-1}%

\usepackage{xr}
\usepackage[pdftex]{graphicx}
\usepackage{amsfonts}
\usepackage{amsmath}
\usepackage{amssymb}
\usepackage{graphicx}
\usepackage{color}
\usepackage{amsmath}
\usepackage{float}
\usepackage{hyperref}

\begin{document}
\title{Closing the gap between atomic-scale lattice deformations and continuum elasticity}

\author{Marco Salvalaglio} \email[\textit{Corresponding Author} \newline \textit{E-mail:} ]{marco.salvalaglio@tu-dresden.de} \thanks{\newline \textit{Address:}  Willersbau (B-209), Zellescher Weg 12-14, 01069 Dresden, Germany. \newline \textit{Tel:}~+4935146335657. \newline \textit{Fax:}~+4935146337096.}
\affiliation{Institute  of Scientific Computing,  Technische  Universit\"at  Dresden,  01062  Dresden,  Germany}
\author{Axel Voigt}
\affiliation{Institute  of Scientific Computing,  Technische  Universit\"at  Dresden,  01062  Dresden,  Germany}
\affiliation{Dresden Center for Computational Materials Science (DCMS), TU Dresden, 01062 Dresden, Germany}
\author{Ken R. Elder}
\affiliation{Department of Physics, Oakland University, Rochester, 48309 Michigan, USA.}

\begin{abstract}
Crystal lattice deformations can be described microscopically by explicitly accounting for the position of atoms or macroscopically by continuum elasticity. In this work, we report on the description of continuous elastic fields derived from an atomistic representation of crystalline structures that also include features typical of the microscopic scale. Analytic expressions for strain components are obtained from the complex amplitudes of the Fourier modes representing periodic lattice positions, which can be generally provided by atomistic modeling or experiments. The magnitude and phase of these amplitudes, together with the continuous description of strains, are able to characterize crystal rotations, lattice deformations, and dislocations. Moreover, combined with the so-called amplitude expansion of the phase-field crystal model, they provide a suitable tool for bridging microscopic to macroscopic scales. This study enables the in-depth analysis of elasticity effects for macro- and mesoscale systems taking microscopic details into account.
\end{abstract}

\maketitle

\section*{Introduction}
Strains and defect-induced deformations have tremendous effects on
the macroscopic properties of single and poly-crystalline materials
\cite{Sethna2017}. These effects have fostered a huge variety of 
studies for more than a century, starting with the first theories describing
the elastic field generated by dislocations in solids \cite{Volterra1907,Taylor1934}. 

Deformation of crystal lattices, although involving changes in the positions of
atoms, are crucial for understanding the behavior of systems defined
on larger length scales \cite{Rollett2015}. Continuum mechanics, and associated continuous elastic fields,
are very useful for describing elastic effects on the mesoscopic and/or macroscopic 
scales. In this approach, a continuous representation of the displacement of atoms in a
lattice with respect to a reference crystal is employed \cite{landau1986}. It
is useful for either relatively simple distortions, as the one given by pure elastic
deformations and rotation, or for deformations induced by the presence of
dislocations \cite{landau1986,Lazar2005,Cai2006,anderson2017}. Indeed, it
can be exploited to provide in-depth studies of material properties allowing
for direct comparisons with experiments and/or a-priori predictions, as, e.g,
for plasticity onset in complex heterostructures \cite{Montalenti2014,Isa2016}
or elasticity effects on material transport mechanisms and morphological
evolution \cite{Rovaris2016,Bergamaschini2016b}.

For some applications, however, continuum mechanics is not enough as neglecting
the description of atoms leads to a crucial loss of information. For instance,
this applies to contributions of the dislocation core to the elastic
field \cite{Ehrlacher2016,Braun2018} and, in turn, to dislocation nucleation,
motion, and reaction. In these cases, in order to describe material properties
by elasticity theory, the elastic field must be described within mesoscale
\cite{Bulatov2006,Devincre2008} or atomistic approaches. Typically, severe
restrictions are present for these methods in the description of long
time- and large length-scales. 

An attempt to overcome the timescale limits of atomistic approaches, by
focusing on diffusive timescales, lead to the development of the so-called phase-field crystal
(PFC) model. It focuses on the dimensionless atomic density field difference, $n$, filtering out vibrations on lattice sites
\cite{Elder2002,Elder2004,Emmerich2012}. It provides good descriptions of
elasticity \cite{Heinonen2014} and dislocation dynamics \cite{Berry2014} even
if it usually requires fine spatial discretizations. This latter limitation is
overcome by the complex \textit{amplitude expansion}
(APFC)\cite{Goldenfeld2005,Athreya2006,GoldenfeldJSP2006,SpatschekPRB2010} of the PFC model for
which both long time scales and large length scales can be examined. It
consists of a coarse-grained representation of the density $n$ that is
expressed by the sum of Fourier modes representing specific lattice symmetries. 
The slowly-oscillating complex amplitudes of these modes, $\eta_j$, are 
then the variables used to characterize the crystalline lattice. Real amplitudes, which may be regarded as a special case of the APFC model, have been also considered \cite{Khachaturyan1983,Khachaturyan1996}, delivering long-range order parameters as in the classical phase-field approaches based on atomistic descriptions. They can be used to account for bridging-scale descriptions of elasticity effects by means of additional contributions as, e.g. in the presence of precipitates, alloys or point defects \cite{Chen1991,Bugaev2002,Tewary2004,Varvenne2012,Varvenne2017}. However, they do not directly encode rotational invariance and elasticity associated to the deformations of the crystal lattice.

Although some intrinsic limitations for large deformations and tilts exist \cite{SpatschekPRB2010}, APFC has proved useful in the advanced modeling of materials as illustrated in studies of elasticity effects \cite{SpatschekPRB2010,Heinonen2014}, compositional domains \cite{Geslin2015}, binary alloys
\cite{ElderPRE2010}, dislocation dynamics \cite{Skaugen2018,Skaugen2018b}, morphology and
motion of dislocation networks at grain boundaries \cite{SalvalaglioGB2018}, and
control of material properties
\cite{Choudhary2012,SalvalaglioAPFC2017,Ofori-Opoku2018}. However, the basic
concept of APFC, namely the coarse-graining of an explicit lattice
representation by focusing on the complex coefficients of Fourier modes, can be
readily applied to any atomistic description as obtained, e.g, from theoretical
modeling, atomistic simulations, or experimental imaging. Still, a direct connection to continuum elasticity is missing.

In this work we first show how to exploit the representation delivered by the complex amplitudes expansion of the PFC model to derive expressions for elastic field components independent of lattice symmetry and system dimensionality. In practice, we describe how to reconstruct strain and rotation fields from the atomic density provided that the complex amplitudes functions of the corresponding Fourier modes are known. Then, we consider numerical simulations for some generic systems involving strained/tilted crystals and apply the new framework in order to depict and analyse the resulting deformations. To this purpose, we numerically solve the equations of the APFC model directly delivering the amplitudes function. Standard simulations as well as simulations which extend the current state of the art for APFC and PFC approaches are presented and discussed. The combination of using a coarse-grained approach as APFC with the detailed analysis of deformations proposed here results in a bridging-scale framework enabling the study of elasticity effects from the micro- to macro-scale while describing microstructural evolution, at variance with many other methods focused only on some of these aspects at once.  Moreover, using the APFC model allows us for its further assessment as a reliable coarse-grained method accounting for microscopic effects. The elastic field in presence of defects, derived from complex amplitudes as computed by APFC simulations, reproduce predictions of continuum mechanics as expected from PFC-based modeling \cite{Elder2002,Elder2004,SpatschekPRB2010}.  Moreover, the results also indicate that APFC includes some deviation from continuum mechanics that may be ascribed to atomistic structure at the dislocation cores, as it has also been observed and discussed in other continuum or atomistic approaches \cite{Lazar2017,Ehrlacher2016,Braun2018}.

\section*{Results and Discussions}

\subsection*{Strain and rotation fields from complex amplitudes}
\label{sec:strain}

The complex amplitude functions, $\eta_j$, entering the APFC model (see \textit{Methods} section) are connected to the deformation of a crystal $\mathbf{u}$ with respect to a reference lattice by the following equation:
\begin{equation}
\eta_j=\phi_j \text{exp}\left(i \mathbf{k}_j \cdot \mathbf{u} \right),
\label{eq:etafromu}
\end{equation}
with $\{\mathbf{k}_j\}$ a specific set of reciprocal space vectors describing the lattice of an undeformed crystal.
Eq.~\eqref{eq:etafromu} defines $N$ independent equations. $\phi_j$'s are the real values
corresponding to the amplitudes in a relaxed, unrotated crystal. They can be
computed by the minimization of the energy functional
in Eq.~\eqref{eq:energyamplitude} assuming constant, real amplitudes for each
different length of $\mathbf{k}_j$. The quantity $A^2
\equiv 2\sum_{j=1}^N |\eta_j|^2$ delivers an order parameter which is constant in the solid/ordered phase, decreases at defects and interfaces, and vanishes when approaching the liquid/disordered phase. Eq.~\eqref{eq:etafromu} can be rewritten as
\begin{equation}
\varphi_j= \mathbf{k}_j \cdot \mathbf{u} ,
\label{eq:phifromu} 
\end{equation}
with $\varphi_j=\arg({\eta_j})=\arctan \left[ \text{Im} (\eta_j)/ \text{Re}
(\eta_j)  \right]$. In order to determine the components of the deformation field
$u_i$ from amplitudes (with $i$=$x,y$ in 2D and $i$=$x,y,z$ in 3D), Eq.~\eqref{eq:phifromu} must be inverted, resulting in a system of $d$ equations with $d$ the dimensionality of the system.  In 2D, by
selecting $d=2$ amplitudes labelled by generic indexes $l$ and $m$, $u_i\equiv
u_i^{\rm 2D}$ results in
\begin{equation}
u_i^{\rm 2D} = \frac{\epsilon_{ij}}{|\mathbf{k}_l \times \mathbf{k}_m|} \left[ k_m^j \varphi_l  - k_l^j \varphi_m \right], %\\
\label{eq:disp2D}
\end{equation} 
with the two components of the displacement field obtained by index
permutations on the group $(i,j)=(x,y)$ and $\epsilon_{ij}$ the 2D Levi-Civita symbol. In 3D, by selecting $d=3$ amplitudes
labelled by generic indexes $l$, $m$, and $n$, $u_i\equiv u_i^{\rm 3D}$ and we obtain
\begin{equation}
\begin{split}
u_i^{\rm 3D} =& \frac{1}{ \mathbf{k}_n \cdot \left( \mathbf{k}_m \times
\mathbf{k}_l \right) }  \left[ \varphi_l \left( k_m^\kappa k_n^j - k_m^j k_n^\kappa
\right) 
\right. \\ 
& \left. + \varphi_m \left( k_n^\kappa k_l^j - k_n^j k_l^\kappa \right) + \varphi_n
\left( k_l^\kappa k_m^j - k_l^j k_m^\kappa \right) \right], %\\
\end{split}
\label{eq:disp3D}
\end{equation}
with the three components of the displacement field obtained by index
permutations on the group $(i,j,\kappa)=(x,y,z)$. Amplitudes must be chosen in order to
have a non vanishing denominator of the prefactor entering Eq.~\eqref{eq:disp3D}.
Without loss of generality we fix here $l=1$, $m=2$ and $n=3$, referring to $\mathbf{k}_j$ with the same length for each symmetry considered here. For small
deformations, the strain tensor $\boldsymbol{\varepsilon}$
can be written
$\boldsymbol{\varepsilon}=(1/2)[\nabla \mathbf{u}+(\nabla \mathbf{u})^{\rm
T}]$. The strain components can then be explicitly computed from $\eta_j$ by means
of spatial derivatives of Eq.~\eqref{eq:disp2D} ($d=2$) or Eq.~\eqref{eq:disp3D}
($d=3$). This leads to $\partial \varphi_j/\partial x_i$ terms. Notice that,
$\varphi_j$ are inherently discontinuous due to their functional form. However,
$\eta_j$ are, by definition, continuous complex functions in both their real
and imaginary part and
\begin{equation} \frac{\partial \varphi_j}{\partial
x_i}= \frac{1}{|\eta_j|^2} \left[ \frac{\partial \text{Im} (\eta_j) }{\partial
x_i}\text{Re} (\eta_j) - \frac{\partial \text{Re} (\eta_j) }{\partial x_i}
\text{Im} (\eta_j) \right] .
\label{eq:derphase}
\end{equation}
Since $|\eta_j|^2=\phi_0^2 > 0$ almost everywhere in the crystal phase, the
terms $\partial \varphi_j/\partial x_i$ and then $\boldsymbol{\varepsilon}$ can be readily computed. $|\eta_j|^2$ only vanishes exactly at the dislocation core position for some indexes, $j$, consistent with continuum elasticity theory.
By exploiting the components of the displacement
field, rotations of the crystal structure with respect to the reference
orientation can also be evaluated as $\boldsymbol{\omega}=\nabla \times \mathbf{u}$.
$\omega_{ij}$ corresponds to the rotational angle in the $x_i$-$x_j$-plane. For
$d=2$, $\omega_{ij} \equiv \omega$, describing the rotation in the
two-dimensional domain. For the sake of readability, the expressions
of $\boldsymbol{\varepsilon}$ and $\boldsymbol{\omega}$ for $d=2,3$, as
functions of $\partial \varphi_j/\partial x_i$, are reported in the
\textit{Supplementary Information}~S2. The resulting deformation fields deliver descriptions similar to advanced continuum theories, as e.g. in Refs.~\cite{Lazar2013,Po2014,Lazar2017}, directly connected here to atomic arrangements via $\eta_j$ functions.

\subsection*{Deformations induced by dislocations}
\label{sec:edgedislo}

Let us consider pairs of dislocations in a 2D triangular lattice that form at 
the interface between layers of different 
atomic spacing in order to accommodate a misfit strain. As performed in
Ref.~\cite{SalvalaglioAPFC2017}, the dislocations can be described by APFC by setting an
initial condition for $\eta_j$ in order to reproduce opposite deformations
(namely initial strains) $\pm\, \varepsilon$. This can be done using 
Eq.~\eqref{eq:etafromu} with an in-plane displacement field $\mathbf{u}(\mathbf{r})$
defined by $u_x=\pm (a_{\rm tri}/L_x) x$ and $u_y=0$, where $\mathbf{r}=x\hat{\mathbf{x}}+y\hat{\mathbf{y}}$ ($+z\hat{\mathbf{z}}$ in 3D). 
\begin{figure}[b]
\includegraphics[width=\linewidth]{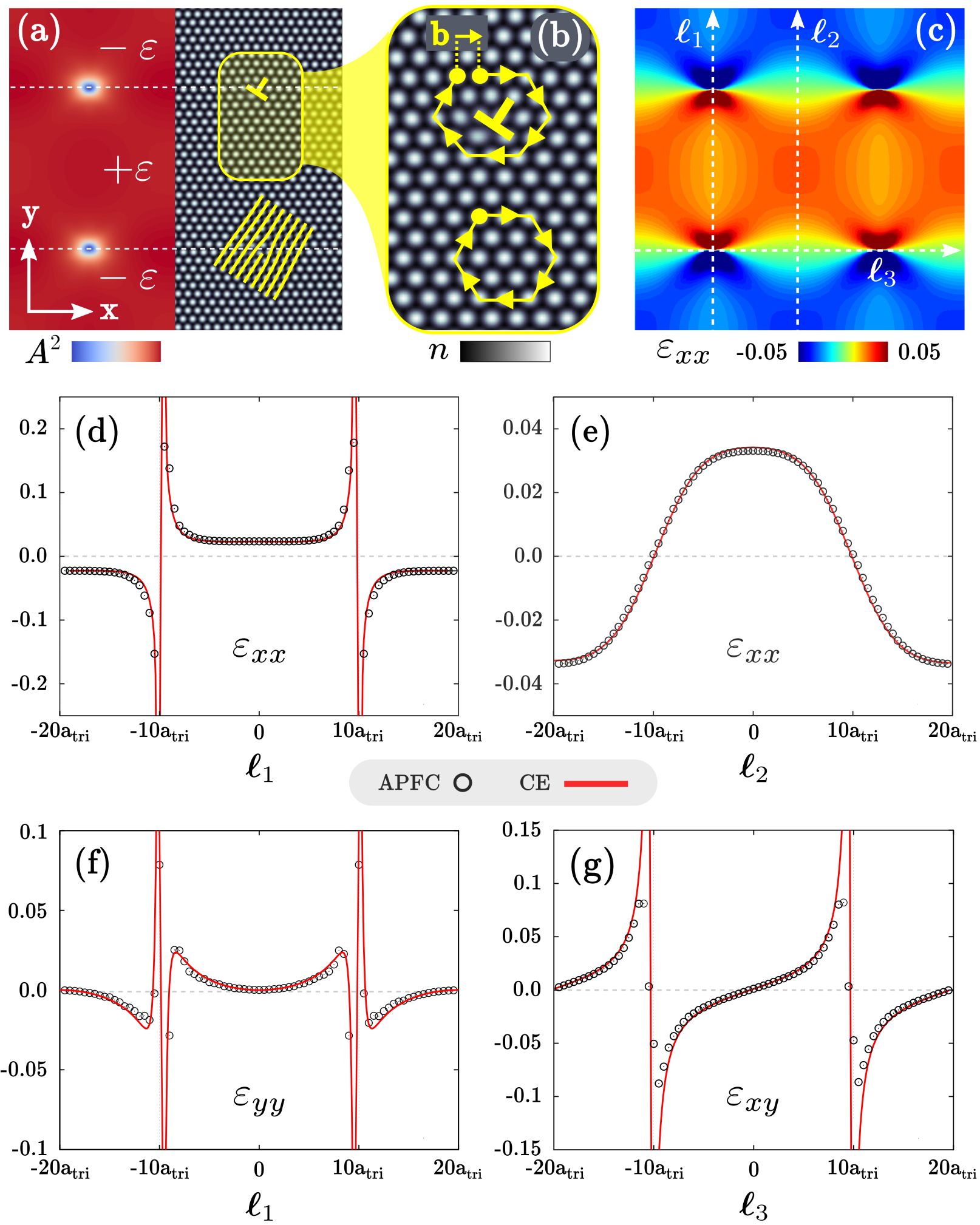} 
\caption{Dislocations and strain field in 2D for a crystal having triangular symmetry. (a) $A^2$ (left),
and $n$ (right), showing a schematics of the position and
orientation of defects as well as atomic planes at and close to one defect. (b) Magnification of $n$ and Burgers vector. (c) $\varepsilon_{xx}$ as computed from $\eta_j$. Comparison between results
from APFC (black circles) and from continuum elasticity (CE) (solid red line)
along specific directions as in panel (c) are shown: (d)
$\varepsilon_{xx}$ along $\mathbf{\ell}_1$, (e) $\varepsilon_{xx}$ along
$\mathbf{\ell}_2$, (f) $\varepsilon_{yy}$ along $\mathbf{\ell}_1$, (g)
$\varepsilon_{xy}$ along $\mathbf{\ell}_3$.}
    \label{fig:figure1}
\end{figure} 
$a_{\rm tri}$ is the distance between
maxima of the density as in Eq.~\eqref{eq:density} for triangular symmetry and $L_x$ is the size of the
computational domain, both evaluated along the $x$ axis. These opposite
displacements are imposed as illustrated in Fig.~\ref{fig:figure1}(a).

The relaxation of such an initial condition until the defect shape is stationary
in a square system of linear dimension $L_i=40a_{\rm tri}$ is illustrated in Fig. \ref{fig:figure1}(a) (details about simulations are reported in the \textit{Methods} section and references therein). 
$A^2$ is shown on the left and the reconstructed density $n$ from Eq.~\eqref{eq:density} 
is shown on the right. Two pairs of dislocations form within the
simulation domain where $A^2$ is constant in the solid and decreases at
defects. Although the APFC approach does not allow for the exact representation
of atoms at the dislocation-core, the lattice distortion is well described as
shown by the illustration of atomic planes by solid lines, highlighting the
presence of one specific defect. The deformation of the lattice can be
quantified at the atomic level by evaluating the Burgers vector, $\mathbf{b}$,
as shown in Fig.~\ref{fig:figure1}(b). Notice that it corresponds to a lattice
spacing in $\hat{\mathbf{x}}$ direction, i.e. $|\mathbf{b}|=4\pi/\sqrt{3}$. $\varepsilon_{xx}$ is shown in
Fig.~\ref{fig:figure1}(c) while, for the sake of completeness, other components
are shown in the \textit{Supplementary Information}~S3.

\begin{figure}[h]
\center
\includegraphics[width=\linewidth]{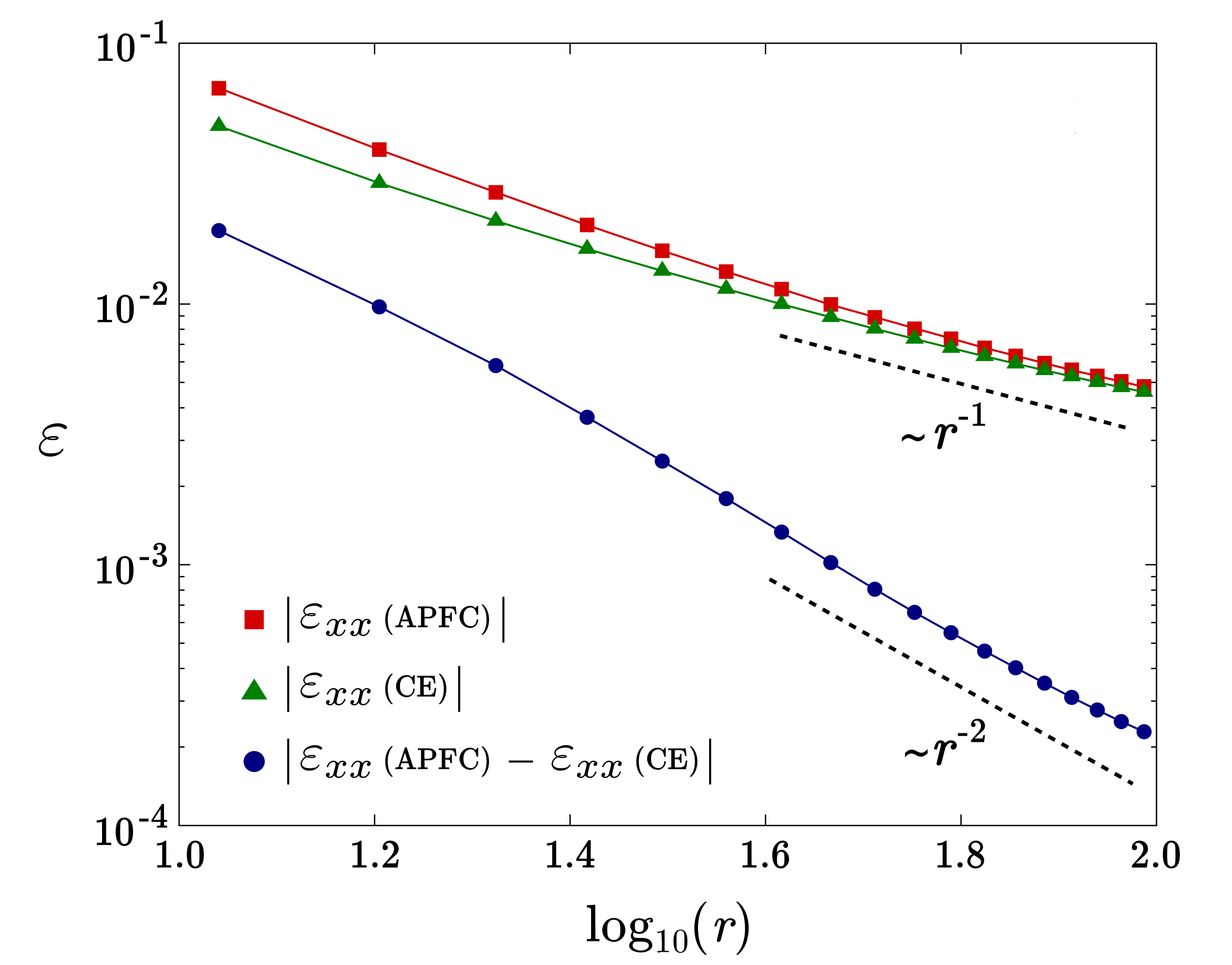} 
    \caption{Scaling of $\varepsilon_{xx}$ along the direction $\ell_1$ as in
Fig.~\ref{fig:figure1}(c) with $r=0$ the position of a dislocation. The value
computed by APFC (red squared) and by continuum elasticity
(green triangles) are shown along with their difference (blue circles). 
    }
    \label{fig:figure2}
\end{figure}
\begin{figure*}
\center
\includegraphics[width=\linewidth]{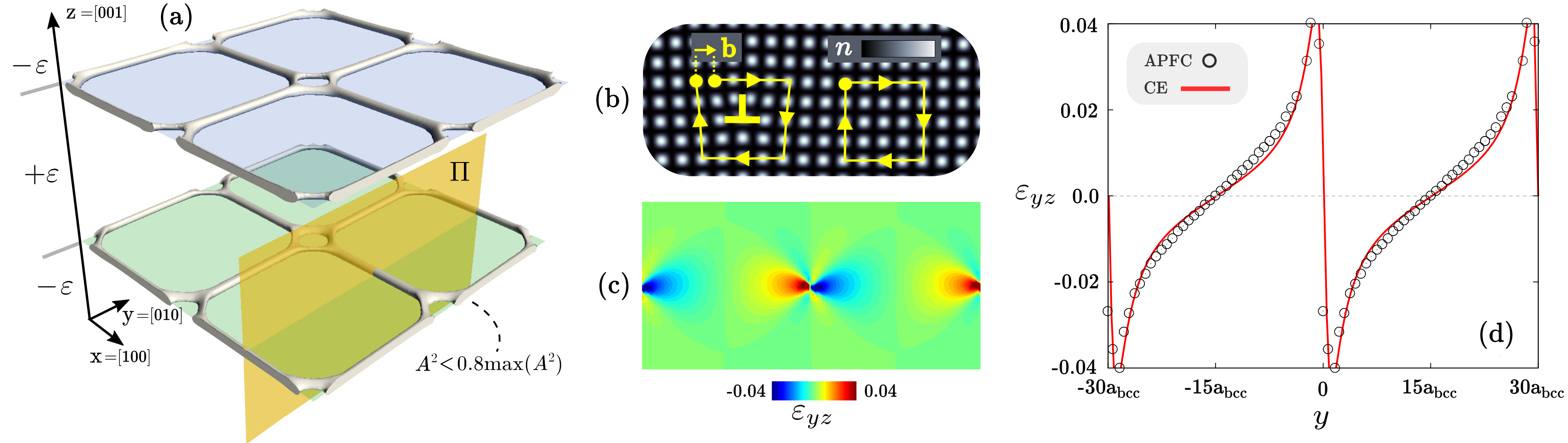} 
    \caption{Dislocations and strain field in 3D for a crystal having bcc
lattice symmetry. (a) Defect network from APFC simulations for a multilayer
configuration. $\Pi$ is the (orange) yz-plane on which next panels focus. (b) reconstructed density at a defect shown in panel (a) and Burgers vector. (c) $\varepsilon_{yz}$ on $\Pi$. (d) Comparison between $\varepsilon_{yz}$ computed from APFC (black circles) and from continuum elasticity (CE) (solid red line)
along the the line parallel to the $\hat{\mathbf{y}}$ direction connecting the defects on $\Pi$.} \label{fig:figure3}
\end{figure*}

The elastic field as computed from the complex amplitudes matches very well with 
continuum elasticity as shown in
Fig.~\ref{fig:figure1}(d)-(g). The latter can be computed provided that the
dislocation character, namely edge, screw or mixed \cite{anderson2017}, as well
as $\mathbf{b}$ and elastic constants are known. Within the APFC model the elastic constants are determined by the parameters entering the energy (see \textit{Methods} section and Eq.~\eqref{eq:energyamplitude}) as discussed in Ref.~\cite{Heinonen2014}. They do not appear explicitly in the equations reported above but affect amplitude values and their spatial derivatives. As reported in the \textit{Supplementary Information}~S4, the defects in Fig.~\ref{fig:figure1}(c) can be modeled as a 2D array of edge dislocations and the strain field according to continuum
elasticity (CE) $\boldsymbol{\varepsilon}^{\rm ce}$ can be computed as
superposition of the elastic field of single dislocations. The elastic constant entering this equations is the Poisson ratio $\nu$ set here to 1/3 as in \cite{Elder2002,Elder2004} (see also \textit{Supplementary Information}~S4). Figs.~\ref{fig:figure1}(d)-(g) show  $\boldsymbol{\varepsilon}^{\rm ce}$ by
solid red lines along $\ell_1$, $\ell_2$, and $\ell_3$ as defined in
Fig.~\ref{fig:figure1}(c). The strain components computed from $\eta_j$'s are
shown by black circles. An almost perfect agreement is found far away from
dislocation cores as observed along $\ell_2$ for $\varepsilon_{xx}$. The same
holds true for the other cases except for regions close to the dislocation cores,
namely at $\ell_i \sim \pm 10 a_{\rm tri}$ where, however, a continuous description of lattice deformations is not
well-posed.

Deeper insight can be
obtained by focusing on the analysis of the elastic field as in
Refs.~\cite{Ehrlacher2016,Braun2018}. Therein, the components of the strain
field in the presence of a straight dislocation, $\varepsilon^{\rm d}_{ij}$,
are decomposed as $\varepsilon^{\rm d}_{ij}=\varepsilon^{\rm
ff}_{ij}+\varepsilon^{\rm core}_{ij}$, with $\varepsilon^{\rm ff}_{ij}$ a
far-field predictor of the elastic field and $\varepsilon^{\rm core}_{ij}$ a
correction due to the core effects. A decay $\varepsilon^{\rm ff}_{ij}
\sim r^{-1}$, with $r$ the distance from the dislocation core, is
expected in agreement with continuum elasticity theory. The dislocation core
effects are characterized by a decay $\varepsilon^{\rm core}_{ij} \sim r^{-2}$ independently on the lattice symmetry, except for some high-symmetry nominal position of the core where $\varepsilon^{\rm core}_{ij} \sim r^{p}$ with $p < -2$ \cite{Ehrlacher2016,Braun2018}.
$\varepsilon^{\rm ce}_{ij}$ corresponds to an explicit expression of
$\varepsilon^{\rm ff}_{ij}$. The
contribution of the dislocation core can then be analysed by evaluating
$\varepsilon^{\rm core}_{ij}=\varepsilon^{\rm apfc}_{ij}-\varepsilon^{\rm ce}_{ij}$, with
$\varepsilon^{\rm apfc}_{ij}$ the elastic field components computed from $\eta_j$. The scaling
of these quantities is illustrated in Fig.~\ref{fig:figure2} where a system as
in Fig.~\ref{fig:figure1} is considered with $L_i=320 a_{\rm tri}$. 
$\varepsilon_{xx}^{\rm apfc}$ (red squared) and $\varepsilon_{xx}^{\rm ce}$  (green triangles) 
are shown along the
line $\ell_1$ (see Fig.~\ref{fig:figure1}(c)) with $r=0$ the position of one of
the defects. In the large $r$ limit the strain shows a decay that is 
very close to $\sim r^{-1}$, but some small deviations are observed.
By evaluating the difference, $\varepsilon_{xx}^{\rm apfc}-\varepsilon_{xx}^{\rm{ce}}$, 
(blue circles) we observe a faster decay that 
scales as $\sim r^{q}$ with $q=-2.05 \pm 0.15$ recalling the more
localized correction given by the dislocation core. We can then
conclude that the elastic field description proposed here not only matches the
description of continuum elasticity far from dislocation cores but also
includes a correction that may be ascribed to atomic-scale effects.

The evaluation of the elastic field can be readily provided also in 3D regardless the lattice
symmetry (see also \textit{Supplementary Information}~S2). We consider now a configuration with peculiar 3D features as shown in
Fig.~\ref{fig:figure3}. An initial condition mimicking layers with opposite,
biaxial strain along the $\hat{\mathbf{x}}$ and $\hat{\mathbf{y}}$ direction is
considered. $\eta_j$'s are initialized as in Fig.~\ref{fig:figure1}
with $u_y=u_x$ \cite{SalvalaglioAPFC2017} and, without loss of generality, we
select a crystal having bcc lattice symmetry (by setting $\mathbf{k}_j$
accordingly, see \textit{Supplementary Information}~S1). The size of the domain is set to $L_i=60 a_{\rm bcc}$, with $a_{\rm bcc}$ the lattice constant for bcc arrangements. Interfaces between layers are (001)
planes. APFC simulations account for the formation of dislocation networks at the interface as shown in Fig.~\ref{fig:figure3}(a). In this figure, the
grey structure corresponds to the region where $A^2<0.8\max(A^2)$, i.e. to the
defects as $A^2$ significantly decreases. The initial
interfaces between layers are higlighted by blue (top) and green (bottom)
planes. The atomic structure at the defects as constructed by Eq.~\eqref{eq:density} and the Burgers vector (here $|\mathbf{b}|=2\pi\sqrt{2}$) are illustrated in Fig.~\ref{fig:figure3}(b) in a small 2D region around the defect lying on the
yz-plane $\Pi$ highlighted in Fig.~\ref{fig:figure3}(a). The strain field
computed from amplitudes is illustrated by means of $\varepsilon_{yz}$
in Fig.~\ref{fig:figure3}(c). By accounting
for the parallel dislocations forming along both the in-plane directions, the
elastic field can be approximated by continuum elasticity (CE). A comparison of $\varepsilon_{yz}$ as obtained from amplitudes (APFC), computed by the
simulations illustrated in Fig.~\ref{fig:figure3}(a), and from CE, by adapting the equations of \textit{Supplementary Information}~S2 (accounting then for two sets of dislocations oriented
along $\hat{\mathbf{x}}$ and $\hat{\mathbf{y}}$ directions and having perpendicular
Burgers vector as in Fig.~\ref{fig:figure3}(b)), is reported in
Fig.~\ref{fig:figure3}(d) showing a general agreement similar to
Fig.~\ref{fig:figure1}. Large length-scale decays as in Fig.~\ref{fig:figure2}, not explicitly addressed here, are expected also in this case \cite{Ehrlacher2016,Braun2018}.

\subsection*{Lattice rotations and polycrystalline systems}
\label{sec:multi}

\begin{figure}[ht]
\center
\includegraphics[width=\linewidth]{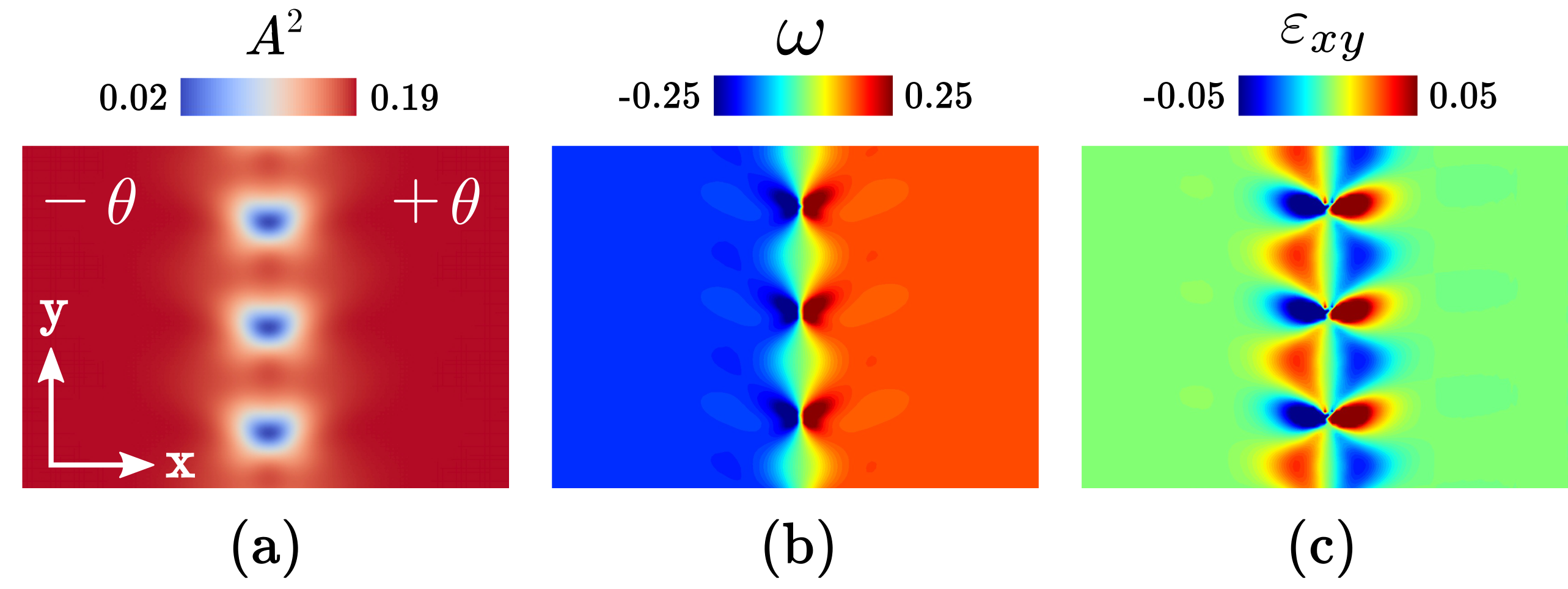} 
    \caption{Deformation and rotation field at a symmetric-tilt GB in 2D. A
(small) portion of the entire simulation domain is reported, illustrating: (a)
$A^2$, (b) $\omega$, (c) $\varepsilon_{xy}$.} \label{fig:figure4}
\end{figure}

We focus here on the analysis of deformations and rotations in polycrystalline
systems, which involve the evolution of small-angle GBs.
We study first a simple system made of a straight GB forming between two crystals with a symmetric tilt. In particular, a rectangular domain, $L_x \times L_y$ with $\hat{\mathbf{x}}=[10]$ and $\hat{\mathbf{y}}=[01]$, is considered with a straight vertical GB at the center. The relative tilt angle between the two crystals, namely $2\theta$, is set by initializing the $\eta_j$ functions as
\begin{equation}
\eta_j=\phi_j \exp \left( i\delta \mathbf{k}_j(\theta) \cdot \mathbf{r} \right),
\label{eq:amprot}
\end{equation}
with $\delta\mathbf{k}_j(\theta)=\mathbf{k}_j \cdot \mathbf{R}(\theta) - \mathbf{k}_j$
and $\mathbf{R}(\theta)$ the counterclockwise rotation matrix. A $\pm \theta$ tilt is
imposed for the left and the right part of the simulation domain respectively
(see also Fig.~\ref{fig:figure4}(a)). By using periodic boundary conditions a GB with infinite extension
is considered. A second GB is also expected, that is shared between the left
and right periodic boundary of the simulation domain. $L_x$ can be chosen
arbitrarily while $L_y$ has to match the periodicity of amplitudes along
$\hat{\mathbf{y}}$. Specific details about this simulation setup can be found in
Ref.~\cite{SalvalaglioAPFC2017}.

Fig.~\ref{fig:figure4}(a) illustrates the result of 
relaxing the aforementioned
initial condition by means of the $A^2$ field in a small region at the GB,
formed by an array of dislocations, showing three defects as minima of $A^2$.
$\omega$ is reported in Fig.~\ref{fig:figure4}(b). A constant value of this
field is obtained when moving away from the GB, while a modulation at the
defects and at the GB is observed, reproducing the effect of the dislocation on
the local orientation of the crystal lattice. The values at which
$\omega$ saturates in the crystals correspond to $\pm
\theta$ imposed in the initial condition. A measure of the tilt angle is then obtained by
exploiting the deformation field. It should be noted, however, that it is not possible to 
directly obtain the tilt angle by inverting Eqs.~\eqref{eq:amprot} due 
to the phase term. Fig.~\ref{fig:figure4}(c) also illustrates the strain
component $\varepsilon_{xy}$, calculated from amplitudes $\eta_j$ as discussed previously. In this instance a significant
superposition of the strain lobes is obtained, due to the proximity of the
dislocations.

\begin{figure}[h]
\center
\includegraphics[width=\linewidth]{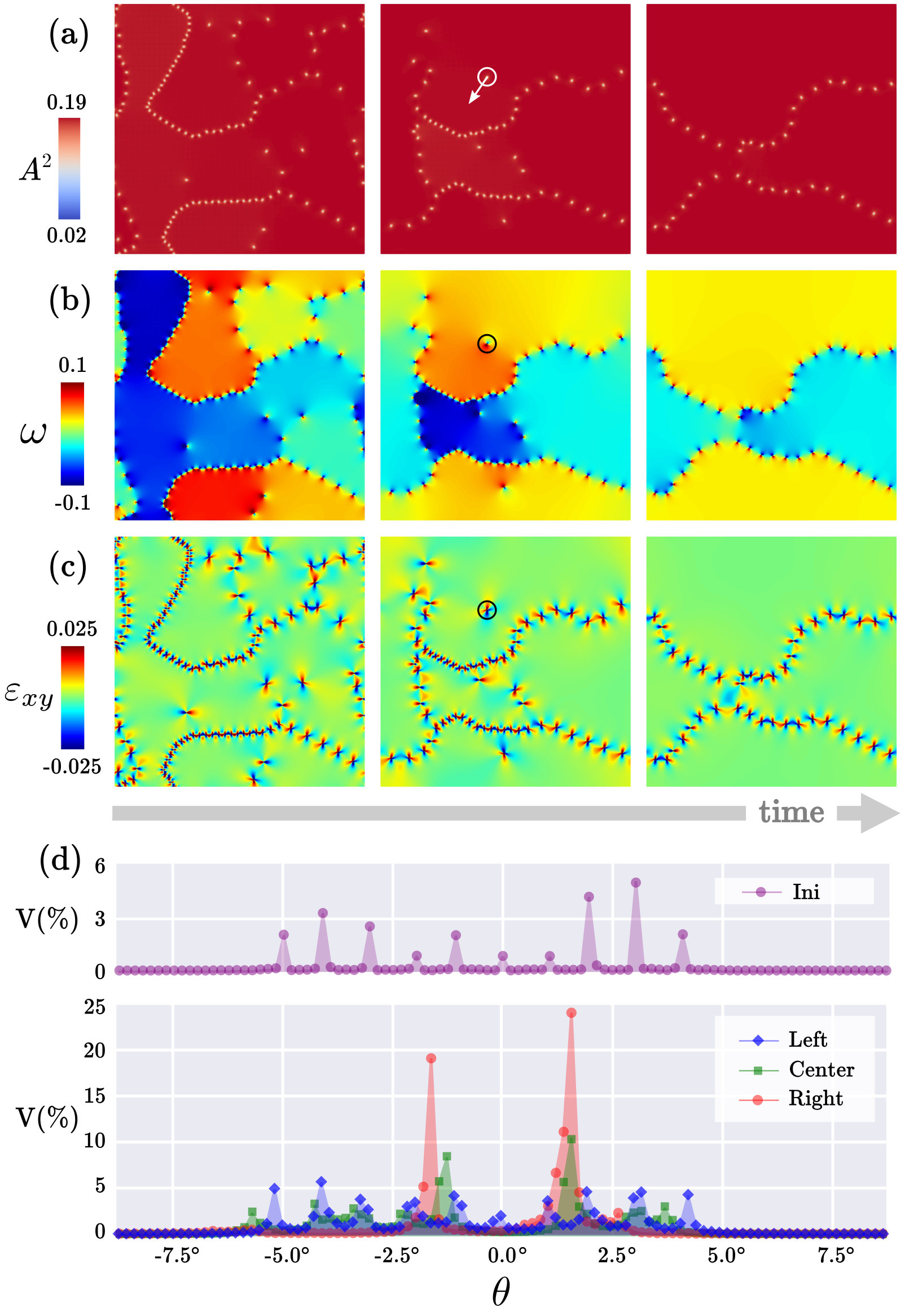} 
    \caption{Deformation and rotation in a polycrystalline system (2D, triangular symmetry). 10 crystals with $\theta \in [-5^\circ,5^\circ)$ evenly spaced, random initial position and radius are considered. The rows illustrate the coarsening dynamics with three stages by means of (a) $A^2$, (b) $\omega$, (c) $\varepsilon_{xy}$. (d) Relative volume of regions having similar tilt angle with uniform binning $\Delta \theta \approx 0.2^\circ$: the initial condition (``Ini'', top panel) and stages reported in (a)-(c), labelled as left, center and right respectively (bottom panel), are shown.}
    \label{fig:figure5}
\end{figure}
\begin{figure*}
\center
\includegraphics[width=\linewidth]{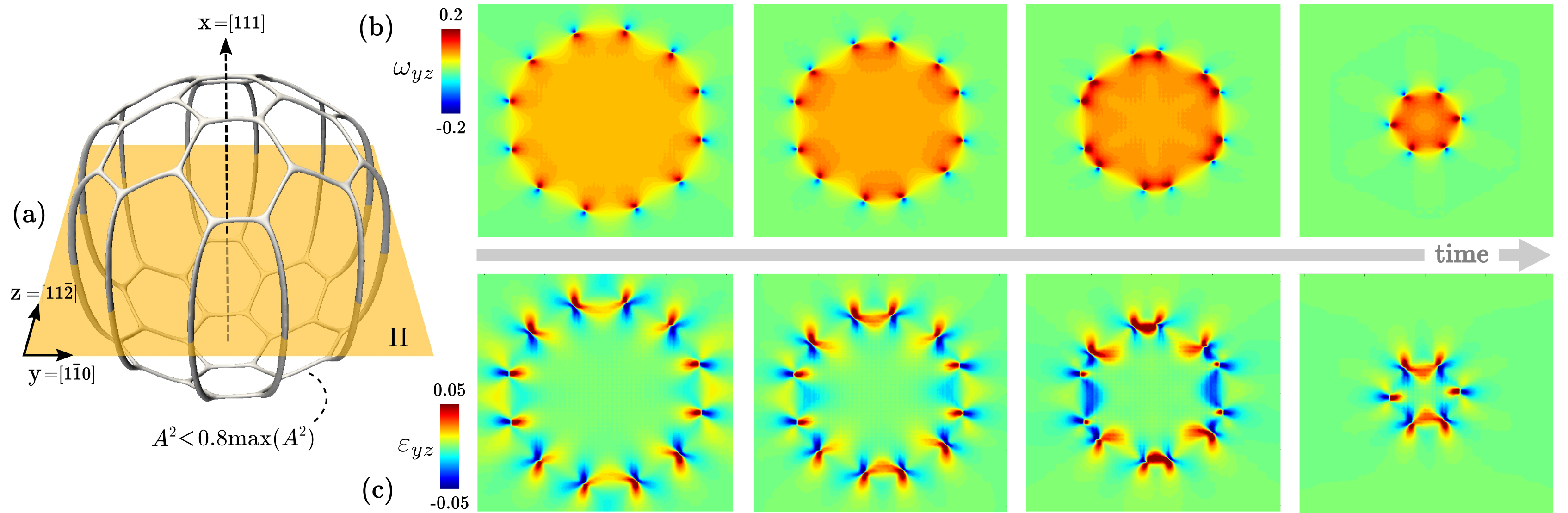} 
    \caption{Deformation and rotation in the presence of a spherical rotated
inclusion with a rotation of $\theta=5^\circ$ about the [111] direction. Fcc
lattice symmetry is considered. (a) Emerging dislocation network at the
interface between the rotated and the unrotated crystal. $\Pi$ is the (orange) yz-plane where the quantities shown in the following panels are evaluated. (b) $\omega_{yz}$. (c) $\varepsilon_{yz}$.} \label{fig:figure6}
\end{figure*}

The importance of extracting the rotation of grains by means of a scalar field
defined everywhere becomes more evident when looking at poly-crystalline systems, as
illustrated in Fig.~\ref{fig:figure5}. Therein a system with $L_i = 320 a_{\rm tri}$ is considered. The initial condition contains 10 crystal 
seeds, namely circular regions with $|\eta_j|>0$ with random positions, random radii 
ranging from $10\pi$ to $20\pi$ and uniformly distributed values of $\theta \in (-5^\circ,5^\circ)$. According to the parameters of the free energy, the crystal phase has the
lowest energy so that the crystals grow and merge forming a complex
network of small-angle GBs made of dislocations. Afterward, these GBs evolve 
resulting in the shrinkage of some grains and the annihilation of dislocations. This is
illustrated in details in Fig.~\ref{fig:figure5} by means of $A^2$, $\omega$ and $\varepsilon_{xy}$ starting from the merged crystal as well as by an analysis of the tilt-angle distribution over time. 
Faceted GBs are obtained (here in 2D corresponding to closed polygonal chains). Moreover, the motion of dislocations leading to the shrinkage of grains occurs along preferred directions related to the crystal lattice and the local tilt, revealing the accurate description achieved by APFC despite its coarse-grained nature. The velocity of a specific defect (within the white and black circles) is highlighted in Fig.~\ref{fig:figure5}. $\omega$ is nearly constant within the grains and varies at the GBs, thus it can be used to identify single grains as they are characterized by different tilts.  Moreover, $\omega$ accounts for the contribution of single dislocations. Indeed, the features of the extracted rotation field allows for the analysis of grains as reported in Fig.~\ref{fig:figure5}(d). Here, the relative volume of regions having the same tilt angle within bins of $\Delta\theta \approx 0.2^\circ$ is shown. The top panel illustrates the initial condition where the crystal seeds cover just a portion of the entire system. The panel at the bottom shows the analysis of the three stages shown in Fig.~\ref{fig:figure5}(a)-(c). By this analysis important information can be extracted, such as the volume fraction (V) occupied by grains in a specific orientation. Peaks in $V$ correspond to the grains with different orientations as shown in Fig.~\ref{fig:figure5}(d). Their broadening is related to the presence of defects, missing in the first stages as the crystals are separated with no dislocations. During the evolution, the relative volume of grains changes and some of them eventually disappear. Notice that before vanishing some peaks shift to larger tilts pointing out a rotation due to the proximity of the dislocations. This is in agreement with 2D atomistic calculations showing an increase of the interface energy and a rotation of the grains during similar processes \cite{Cahn04,Wu2012,Heinonen2014}.

\begin{figure*}
\center
\includegraphics[width=\linewidth]{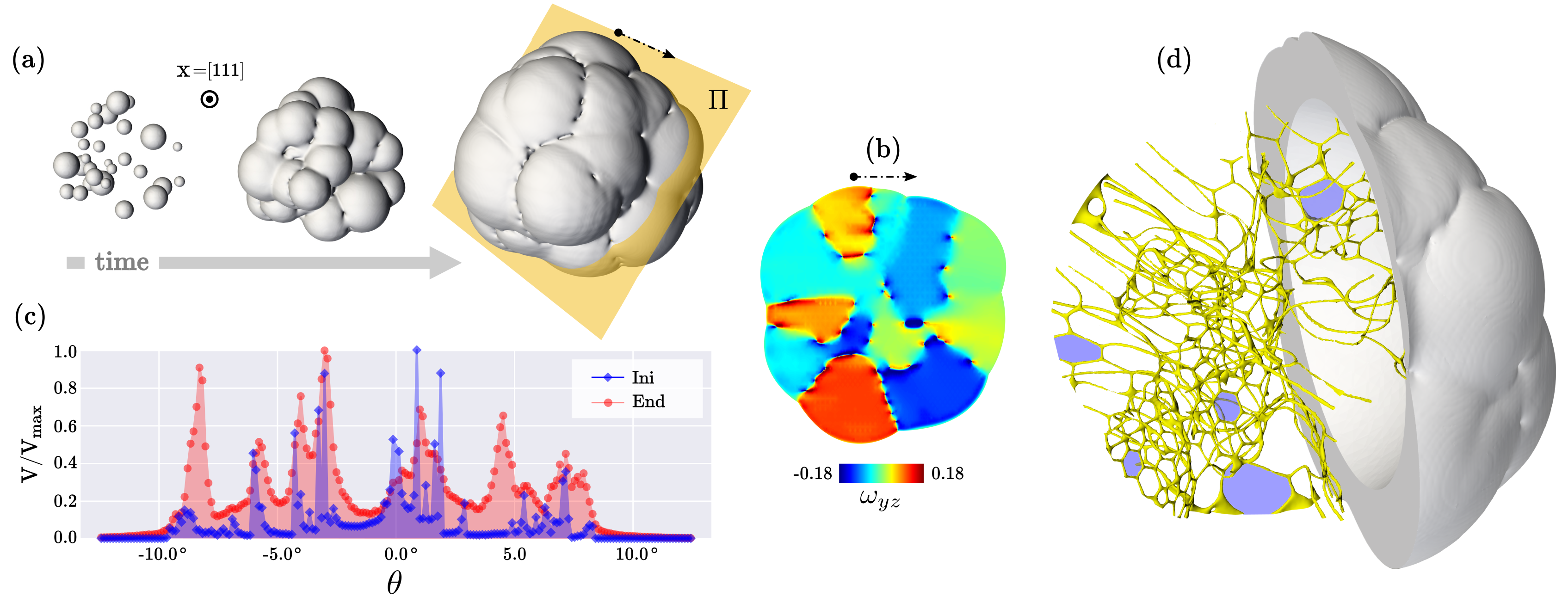} 
    \caption{Growth of randomly tilted crystal seeds (3D, fcc lattice symmetry). 30 crystals with random tilt $\theta \in (-10^\circ,10^\circ)$ about the [111] direction are considered with random initial position and radius. (a) Three stages during the growth of the polycrystalline system. (b) $\omega_{yz}$ in the plane $\Pi$ illustrated in panel (a). (c) Normalized volume fraction of crystal with similar rotation of the first (``Ini'') and last (``End'') stages in panel (a) within bins of $\Delta \theta=0.1^\circ$. (d) Defects (yellow network) within a spherical region at the center of the growing polycrystal (last stage in panel (a)).} \label{fig:figure7}
\end{figure*}

A three-dimensional system involving rotations and consisting of a single spherical crystal that is 
rotated with respect to the larger single crystal surrounding it is  
shown in Fig.~\ref{fig:figure6}.  $L_i=50a_{\rm fcc}$ with $a_{\rm fcc}$ the lattice constant for fcc arrangements.
A rotation of $5^\circ$ about the $[111]$ direction is set. The initial condition for
$\eta_j$ is set by exploiting Eq.~\eqref{eq:amprot} (see also
Ref.~\cite{SalvalaglioGB2018}). The resulting dislocation network after a first relaxation phase
is shown in Fig.~\ref{fig:figure6}(a). As in Fig.~\ref{fig:figure3}(a), the grey structure corresponds to the dislocations. A closed network of defects forms, having
the peculiar structure of defects at twisted GBs when the surface normal
of the spherical inclusion approaches the rotational axis. As expected by classical grain growth theory \cite{Doherty1997} and 3D simulations \cite{SalvalaglioGB2018} this defect network evolves leading to the shrinking of the rotated inclusion. As noticed first in Ref.~\cite{Yamanaka2017}, where PFC simulations of a similar configuration with bcc lattice symmetry are reported, an anisotropic shrinkage of the grain occurs, that is faster along the rotational axis as a result of the dynamics and reaction of dislocations. However, an almost linear scaling of the grain boundary extension is expected \cite{Yamanaka2017,SalvalaglioGB2018}. The extraction of the rotation field on the yz-plane $\Pi$ highlighted in Fig.~\ref{fig:figure6}(a) is illustrated in
Fig.~\ref{fig:figure6}(b) at different stages during the evolution.
$\omega_{yz}$ is zero outside the inclusion and almost constant
inside. Similarly to Fig.~\ref{fig:figure5}(a), a short
wavelength modulation is observed at defects. From the $\omega_{yz}$ color map, a rotation of the grain is observed \cite{Cahn04,Wu2012,Heinonen2014,Yamanaka2017}. The corresponding $\varepsilon_{yz}$ field is shown in  Fig.~\ref{fig:figure6}(c). 

A 3D ploycrystalline system is shown in Fig.~\ref{fig:figure7}, where the growth of 30 crystal seeds having fcc lattice symmetry with random rotation $\theta \in (-10^\circ,10^\circ)$ about the [111] direction, random position and size is simulated. $L_i=100a_{\rm fcc}$ while the seeds are generated within a distance $40a_{\rm fcc}$ from the center. The evolution is illustrated in Fig.~\ref{fig:figure7}(a) by means of three snapshots showing the region where $A^2$ is significantly larger than zero. The rotation field in the $\Pi$ plane is shown in Fig.~\ref{fig:figure7}(b). Fig.~\ref{fig:figure7}(c) illustrates the analysis of the local rotations in terms of the volume of regions with the same angles within bins of $\Delta \theta = 0.1^\circ$ normalized with respect to the larger one. It delivers similar features as discussed Fig.~\ref{fig:figure5}(b). In particular, the initial configuration exhibits smaller angular dispersion than later stages, while peaks broaden due to the formation of dislocations. The capability of the adopted framework can be further appreciated in Fig.~\ref{fig:figure7}(d), where the complex dislocation network forming in a central (spherical) region of the growing polycrystal (last step of Fig.~\ref{fig:figure7}(a)) is shown. Features of the dislocation network observed in Fig.~\ref{fig:figure6}(a) can be here recognized, as hexagonal arrangements and elongated defects according to direction of the normal of the interface between grains. Similar dislocation networks with different spacing (e.g. see the presence of hexagons with different sizes at GB highlighted as blue shaded region) are obtained within the polycrystal due to the different relative rotations between grains.

The analysis of polycrystalline systems allows us to summarize the main findings of this work. A continuous description of lattice deformations exploiting the complex amplitudes of
Fourier modes representing the periodicity of crystal lattices has been derived. Deformation fields can then be readily computed from an atomistic representation of crystals, provided that amplitude functions can be extracted, without any \textit{ad-hoc} post-processing procedure and independently of system dimensionality and lattice symmetry. This framework has been shown to achieve its full potential combined with the APFC approach. Other than trivially providing a direct access to amplitudes and allowing for describing either strain or rotated crystals, the APFC model easily allows for large length-scale simulations approaching the ones typical of continuum theories, still retaining essential microscopic features. This has been further demonstrated by showing that APFC encodes dislocation-core contributions to the elastic field, which are missing in standard continuum elasticity and usually requires more refined theories \cite{Lazar2017}. The connection between amplitudes and displacements or strains enables the detailed study of the effect of any deformation which are encoded by continuous deformation fields such as displacements or strain/stress due to single dislocations or due to external loads. We explicilty illustrated that extracting the local orientation as continuous field directly enables microstructural analysis. In addition, it will enable the development of optimized numerical method as, e.g., orientation-based meshing criterion for APFC \cite{PrecondAPFC2019}.

It is worth recalling that the limitation of the APFC model in describing large deformations \cite{SpatschekPRB2010} poses some constraints to the numerical approach adopted here, but not on the derivation of the deformation fields and on the possible analysis of deformations extracted from other atomistic frameworks (e.g. the PFC model itself or experiments). Moreover, some attempts to overcome this APFC limit have been recently proposed \cite{Bercic2018}. Notice that several PFC type models, which are based on a continuous probability density can be easily converted in the proper set of amplitudes, and are therefore naturally compatible with the descriptions illustrated in this work (see e.g. Refs.~\cite{Kocher2015,Wang2016}). In addition, work on extensions of the PFC model can be done in order to derive the corresponding amplitude expansions as done, for instance, in Ref.~\cite{Ofori-Opoku2013} with the so-called XPFC model \cite{Greenwood2010}. 

\section*{Methods}

\subsection*{Amplitude expansion of phase-field crystal model}
\label{sec:model}

The PFC model accounts for the lattice structure by means
of a continuous periodic field $n$ describing the dimensionless atomic probability density
\cite{Elder2002,Elder2004,Emmerich2012}. It is based on a
free-energy functional, $F_n$, that reads
%describes a first-order transition between
%a disordered/liquid phase, where $n$ is constant, and an ordered/crystalline
%phase, where $n$ is periodic. $F_n$ can be written, 
\begin{equation}
F_n=\int_{\Omega} \left[\frac{\Delta B_0}{2}n^2+\frac{B^x_0}{2} n(1+\nabla^2)^2n 
-\frac{t}{3}n^3+\frac{v}{4}n^4 \right]d\mathbf{r},
\label{eq:F_PFC}
\end{equation}
where $\Delta B_0$, $B_0^x$, $v$ and $t$ are parameters as in Ref.~\cite{Elder2007}.
In the crystalline state $n$ can be generally approximated as sum of plane waves as 
\begin{equation}
n(\mathbf{r})=n_0(\mathbf{r})+\sum_{j=1}^N \left[ \eta_j(\mathbf{r}) e^{i\mathbf{k}_j \cdot
\mathbf{r}}+ \eta_j^*(\mathbf{r}) e^{-i\mathbf{k}_j \cdot \mathbf{r}}\right],
\label{eq:density}
\end{equation}
with $n_0(\mathbf{r})$ the average density, $\eta_j(\mathbf{r})$ the amplitude
of each plane wave and $\mathbf{k}_j$ the reciprocal space vector representing
a specific crystal symmetry (see \textit{Supplementary Information}~S1).  In the
so-called amplitude expansion of the PFC model (APFC)
\cite{Goldenfeld2005,Athreya2006,GoldenfeldJSP2006}, these amplitudes are 
the variables used to describe a given crystalline system. Lattice symmetries are described by means of a fixed
set of vectors $\mathbf{k}_j$. Complex amplitudes
functions $\eta_j(\mathbf{x})$ allow for distortions and rotations of the crystal lattice with respect to a reference
state accounted for by $\mathbf{k}_j$ vectors. The free energy, expressed in terms
of $\eta_j$'s, reads  
\begin{equation}
\begin{split}
F=&\int_{\Omega} \bigg[\frac{\Delta B_0}{2}A^2+\frac{3v}{4}A^4 +\sum_{j=1}^N
\left ( B_0^x |\mathcal{G}_j \eta_j|^2-\frac{3v}{2}|\eta_j|^4 \right ) \\
&+f^{\rm s}(\{\eta_j\},\{\eta^*_j\}) \bigg]  d \mathbf{r}, 
\end{split}
\label{eq:energyamplitude}
\end{equation}
with $\mathcal{G}_j\equiv \nabla^2+2i\mathbf{k}_j \cdot \nabla$ and  $A^2
\equiv 2\sum_{j=1}^N |\eta_j|^2$. The term $f^{\rm s}(\{\eta_j\},\{\eta_j^*\})$
corresponds to a complex polynomial of $\eta_j$ and $\eta_j^*$ and depends 
on the specific crystalline symmetry as reported in Ref.~\cite{SalvalaglioAPFC2017} 
for triangular, body-centered cubic (bcc) and face-centered cubic (fcc) lattices. 
The evolution laws for $\eta_j$'s read
\begin{equation}
\frac{\partial \eta_j}{\partial t} =-|\mathbf{k}_j|^2 \frac{\delta F}{\delta \eta_j^*}.
\label{eq:amplitudetime}
\end{equation}

\subsection*{Simulations}
The simulations reported in this work are performed with the aid of High Performance Computing facilities \cite{JURECA} exploiting the finite element toolbox AMDiS \cite{Vey2007,Witkowski2015} with a semi-implicit
integration scheme and mesh adaptivity as reported in Ref.~\cite{SalvalaglioAPFC2017}. Large, three-dimensional simulations as reported in Fig.~\ref{fig:figure7} have been obtained thanks to an improvement of the numerical approach described elsewhere \cite{PrecondAPFC2019}. Periodic boundary conditions are used for all the boundaries of the simulation domains. To describe crystalline phases, the parameters entering the free
energy are set to favor the crystal phase as follows: $B^x=0.98$, $v=1/3$,
$t=1/2$ and $\Delta B=0.02$
\cite{ElderPRE2010,SalvalaglioAPFC2017,SalvalaglioGB2018}. 

%\section*{Data Availability}
%The data and the codes that support the findings of this study are available from the corresponding author upon reasonable request.

\section*{Acknowledgements}
We acknowledge C. Ortner for fruitful discussions about the decay of the elastic field in the presence of dislocations and S. Praetorius for the optimization of numerical simulations.
M.S. acknowledges the support of the Postdoctoral Research Fellowship awarded by the Alexander von Humboldt Foundation. A.V. acknowledges support from the German Research Foundation under Grant No. Vo899/20 within SPP 1959. K.R.E. acknowledges financial support from the National Science Foundation under Grant No. DMR1506634. We acknowledge support by the Open Access Publication Funds of the SLUB/TU Dresden. We also gratefully acknowledge the computing time granted by the John von Neumann Institute for Computing (NIC) and provided on the supercomputer JURECA at J{\"u}lich Supercomputing Centre (JSC), within the Project No. HDR06,  and by the Information Services and High Performance Computing (ZIH) at TU Dresden.

%\section*{Competing Interests}
%The authors declare no competing interests. 

%\section*{Author Contributions}
%M. S. developed the theoretical approach and performed the simulations. M.S., A.V., K. R. E conceived the investigation and wrote the paper. 

\newpage

\begin{figure*}
\centering
\vspace*{-1.0cm}\hspace*{-1.9cm} \includegraphics[page=1,scale=1]{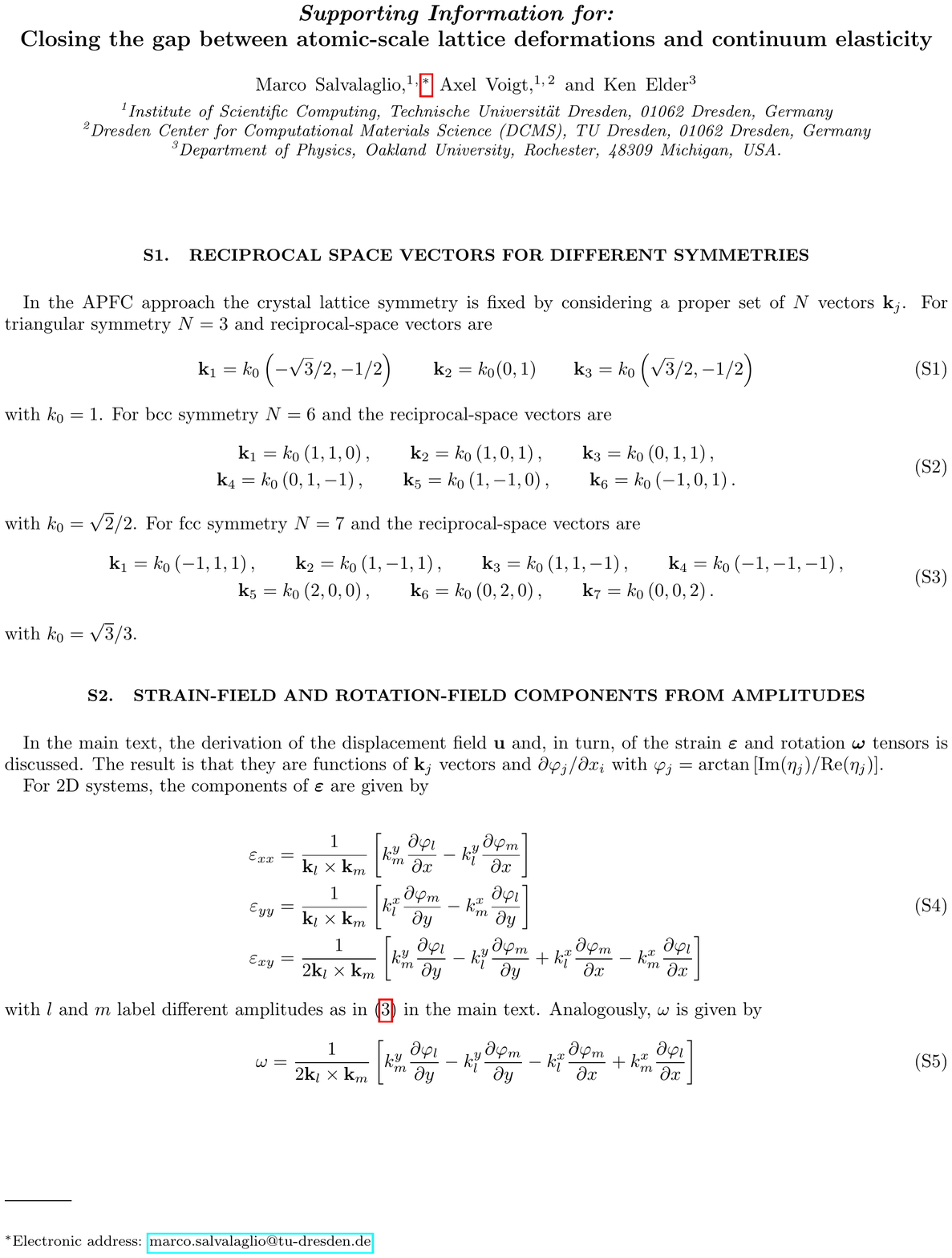}
\end{figure*}

\begin{figure*}
\centering
\vspace*{-1.0cm}\hspace*{-1.9cm} \includegraphics[page=2,scale=1]{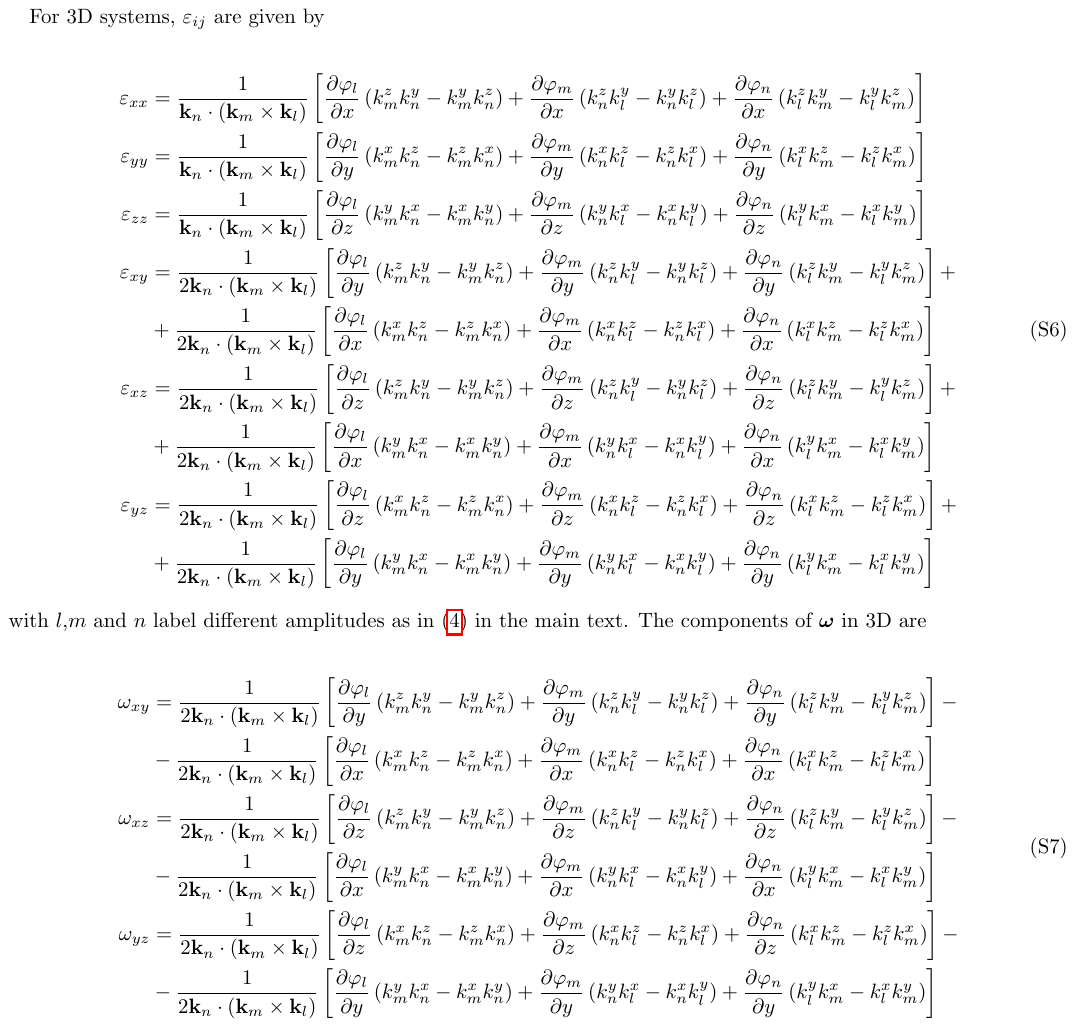}
\end{figure*}

\begin{figure*}
\centering
\vspace*{-1.0cm}\hspace*{-1.9cm} \includegraphics[page=3,scale=1]{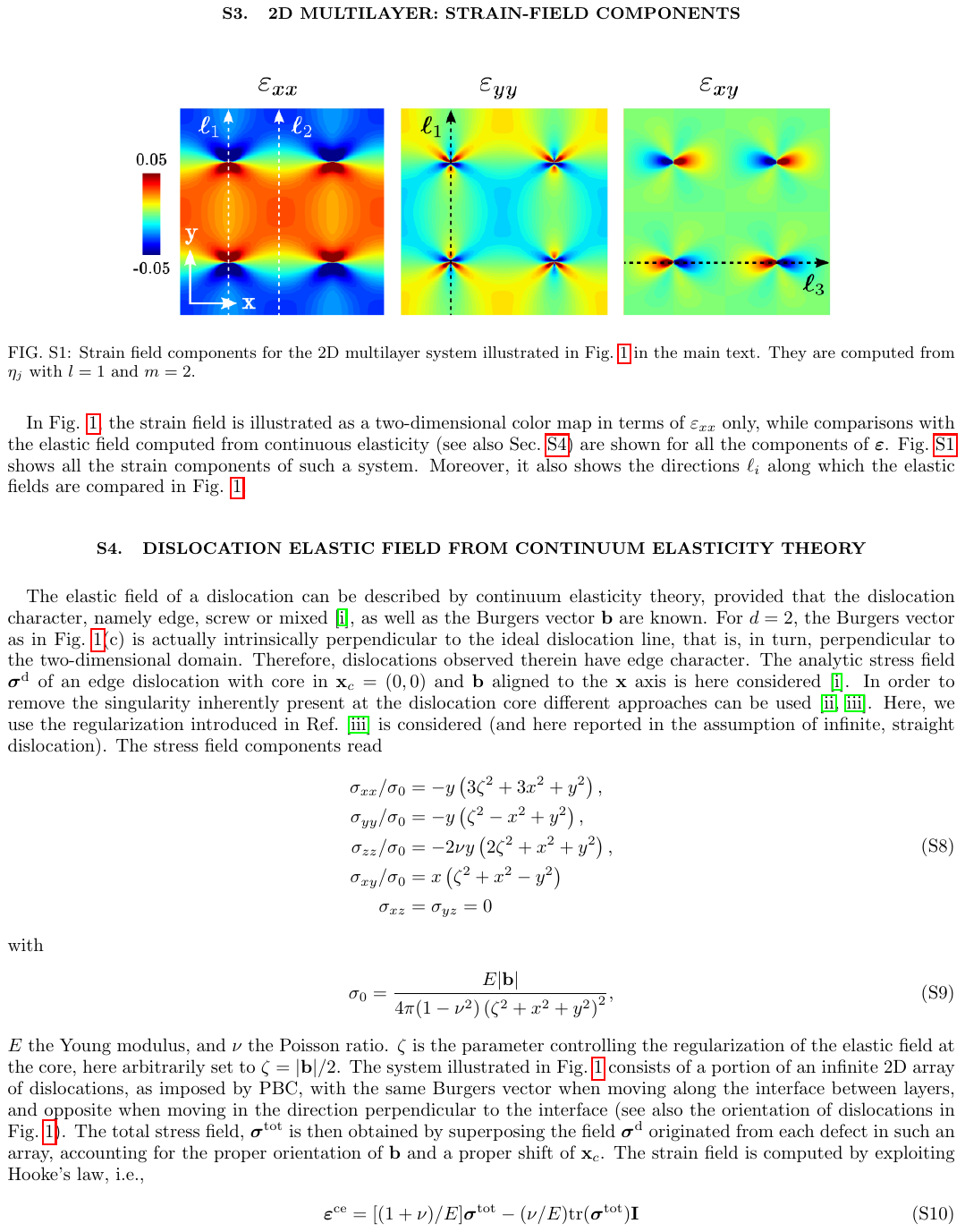}
\end{figure*}

\begin{figure*}
\centering
\vspace*{-1.0cm}\hspace*{-1.9cm} \includegraphics[page=4,scale=1]{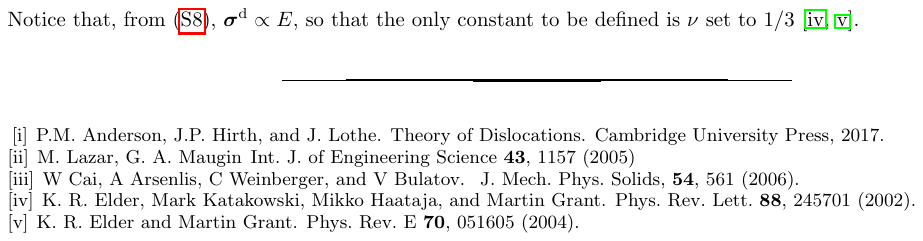}
\end{figure*}

\end{document}